\documentstyle[12pt]{article}
\begin{document}
%

\newskip\humongous \humongous=0pt plus 1000pt minus 1000pt
\def\caja{\mathsurround=0pt}
\def\eqalign#1{\,\vcenter{\openup1\jot \caja
 \ialign{\strut \hfil$\displaystyle{##}$&$
 \displaystyle{{}##}$\hfil\crcr#1\crcr}}\,}
\newif\ifdtup
\def\panorama{\global\dtuptrue \openup1\jot \caja
 \everycr{\noalign{\ifdtup \global\dtupfalse
 \vskip-\lineskiplimit \vskip\normallineskiplimit
 \else \penalty\interdisplaylinepenalty \fi}}}
\def\eqalignno#1{\panorama \tabskip=\humongous
 \halign to\displaywidth{\hfil$\displaystyle{##}$
 \tabskip=0pt&$\displaystyle{{}##}$\hfil
 \tabskip=\humongous&\llap{$##$}\tabskip=0pt
 \crcr#1\crcr}}
\jot = 1.5ex
\def\baselinestretch{1.2}
\parskip 5pt plus 1pt
\catcode`\@=11
\@addtoreset{equation}{section}
\def\theequation{\arabic{section}.\arabic{equation}}
\def\@normalsize{\@setsize\normalsize{15pt}\xiipt\@xiipt
\abovedisplayskip 14pt plus3pt minus3pt%
\belowdisplayskip \abovedisplayskip
\abovedisplayshortskip \z@ plus3pt%
\belowdisplayshortskip 7pt plus3.5pt minus0pt}
\def\small{\@setsize\small{13.6pt}\xipt\@xipt
\abovedisplayskip 13pt plus3pt minus3pt%
\belowdisplayskip \abovedisplayskip
\abovedisplayshortskip \z@ plus3pt%
\belowdisplayshortskip 7pt plus3.5pt minus0pt
\def\@listi{\parsep 4.5pt plus 2pt minus 1pt
     \itemsep \parsep
     \topsep 9pt plus 3pt minus 3pt}}
\relax
\catcode`@=12
\evensidemargin 0.0in
\oddsidemargin 0.0in
\textwidth 6.0in
\textheight 8.5in
\hoffset .7 cm
\voffset -1 cm
\headsep .25in
\catcode`\@=11
\def\section{\@startsection{section}{1}{\z@}{3.5ex plus 1ex minus
   .2ex}{2.3ex plus .2ex}{\large\bf}}

\def\thesection{\arabic{section}}
\def\thesubsection{\arabic{section}.\arabic{subsection}}
\def\thesubsubsection{\arabic{subsubsection}.}
\def\appendix{\setcounter{section}{0}
 \def\thesection{Appendix \Alph{section}}
 \def\theequation{\Alph{section}.\arabic{equation}}}
\newcommand{\beq}{\begin{equation}}
\newcommand{\eeq}{\end{equation}}
\newcommand{\bea}{\begin{eqnarray}}
\newcommand{\eea}{\end{eqnarray}}
\newcommand{\beas}{\begin{eqnarray*}}
\newcommand{\eeas}{\end{eqnarray*}}
\newcommand{\defi}{\stackrel{\rm def}{=}}
\newcommand{\non}{\nonumber}
\def\de{\partial}
\def\si{\sigma}
\def\dim{\hbox{\rm dim}}
\def\sup{\hbox{\rm sup}}
\def\inf{\hbox{\rm inf}}
\def\Arg{\hbox{\rm Arg}}
\def\Im{\hbox{\rm Im}}
\def\Re{\hbox{\rm Re}}
\def\Res{\hbox{\rm Res}}
\def\Max{\hbox{\rm Max}}
\def\Abs{\hbox{\rm Abs}}
\def\infi{\infty}
\def\nrm{\parallel}

\def\1{{\rm 1}}
\def\s{{\sigma}}
\def\e{{\cal E}}
\def\o{{\cal O}}
\def\om{{{\cal O}_m}}
\def\dm{{{\partial}_m}}
\def\fp{{\rm Fp}_\mu}
\def\au{{a_1}}
\def\ad{{a_2}}
\begin{titlepage}
\begin{center}
{\Large
All order I.R. finite expansion for short distance behavior
of massless theories perturbed by a relevant operator.}
\end{center}
\vspace{1ex}
\begin{center}
{\large
Riccardo Guida$^{1}$ and Nicodemo Magnoli$^{1,2}$}
\end{center}
\vspace{1ex}
\begin{center}
{\it $^{1}$ Dipartimento di Fisica -- Universit\`a di Genova\\
     Via Dodecaneso, 33 -- 16146 Genova, Italy}\\
{\it $^{2}$ Istituto Nazionale di Fisica Nucleare- Sez. Genova\\
     Via Dodecaneso, 33 -- 16146 Genova, Italy}
\end{center}
\begin{center}
e-mail: guida@ge.infn.it, magnoli@ge.infn.it
\end{center}
\medskip
{\bf ABSTRACT:}
We consider here  renormalizable theories
without relevant couplings and present an
I.R. consistent technique to study corrections to short distance
behavior
(Wilson O.P.E. coefficients) due to a relevant perturbation.
Our method is the result of a complete reformulation of recent works on the
field, and is characterized by a
 more  orthodox treatment of U.V. divergences that allows
for simpler formulae and  consequently an explicit
all order (regularization invariant) I.R. finitess proof.
Underlying hypotheses are discussed in detail and found to be
satisfied in conformal theories that constitute a natural field of
application of this approach.

\vfill
\begin{flushleft}
GEF-Th-12/1995
 \hfill 11/ 1995
\end{flushleft}

\end{titlepage}

\section{Introduction}\label{introduction}

We present here a consistent technique to study the
corrections to the  short distance behavior
of a  renormalizable
euclidean quantum field theory
without relevant couplings (i.e. with positive mass
dimension), when a relevant perturbation is introduced.

It is well known that  expansions in relevant
parameters around
field theories without a mass scale  are plagued
by  I.R. divergences.
Considerable effort on how to deal with these
singularities has been done,
for generic super-renormalizable
quantum field theories \cite{jt}
as well as for   some  specific cases
\cite{sz},
but   the present situation is such that any general and
(especially) practical
method that could face these I.R. divergences  is always welcome.

The possibility of doing an I.R. finite expansion
in a relevant parameter is particularly appealing in the case
of (two dimensional) conformal field theories.
After the first pioneering work \cite{bpz},
understanding on this  subject
has been greatly increased and many exact results
have been found
 (see for instance \cite{review}). In addition
it soon became  clear that those models  can be taken as
a starting point to get information on non conformally
invariant theories \cite{c}, \cite{campomagnetico}.
{}From a point of view of statistical mechanics, \cite{review2},
conformal field theories are fixed points
of the renormalization group and the line of research
we are considering can be rephrased as
"how to go outside the fixed point".
We think that  the
technique introduced here
 finds a natural field
of application  within this framework
because
it  gives  a {\it general} way to describe the
short distance behavior of a theory  outside
the fixed point, useful to test eventual exactly solved deformations
but essential to give new light in all the other cases.

The method described in this paper
 is a reformulation  of the independent works
of Al.B. Zamolodchikov \cite{zamo} and of
 Mikhak and  Zarkesh \cite{mz} (that applied techniques developed by
Sonoda \cite{sonoda1}).

The first idea underlying the  approach is present already in
earlier papers  of  Wilson \cite{wilson} and Wegner \cite{wegner}
(see also the more recent \cite{sonoda2}):
to deal with I.R. singularities
is essential to redefine composite
fields in such a way to confine
the terms non analytic in the generalized masses
(potentially present  in all the interesting
  correlators)
  inside the V.E.V. of the composite fields,  enforcing
in this way the regularity
of
Wilson coefficients.

The second step is to  give a formal
 expression of the derivatives of Wilson coefficients with
respect to the generalized masses in terms of (generally unknown)
correlators of the
deformed theory:
this has been realized by   perturbative expansions in \cite{zamo},
and by using  the "Variational Formulae" in
\cite{sonoda1} (a kind of Action Principle, see Section \ref{uv}).

The third step is to regularize the theory with an I.R. cutoff
 and to take  the limit of zero generalized masses: clearly at this stage
one has to face the potential I.R. divergences.
The other  main idea  of the approach, \cite{zamo,sonoda1}
(and \cite{mz} for an explicit proof
up to second order),
 is  that the   convergence properties  of O.P.E.
should  guarantee the cancellation of I.R. divergences.
After keeping into account this fact
 the
derivatives of the Wilson coefficients
can be computed by evaluation of I.R. cutoff
integrals involving  exactly known correlators of the massless
 field theory.

Our present contribution is an improvement and an extension
of the previously mentioned   works,
in order to give a simplified  and U.V.
consistent method with well defined applicability conditions
that is I.R. finite at all orders.

First in Section \ref{ipotesi}
we state { clearly} which are the underlying, minimal hypotheses
to be satisfied for the method to work.

Secondly, by using the orthodox renormalized Action Principle
 (that has
 a simpler form with respect to \cite{sonoda1}),
 an explicit
inductive proof of I.R. finitess
and a detailed discussion of
the implications of cancellation of I.R.
divergences is given
 in Section \ref{ir}.
It is also shown that final expressions do not depend on
the  concrete form  of the I.R. regularization (whose choice becomes a
matter of convenience).

In Section \ref{uv} we  treat the
correct definition  of renormalized composite operators
 to be chosen in such a way that Action Principle holds.
We  shall see that from this requirement
arises the problem of existence of numerical constants (essentially
present in the operators' V.E.V.)
whose expression in terms of
deformed theory correlators is {\it well defined} but in general
uncomputable without additional  information.

In Section \ref{toy} we compute  the mass corrections to the short distance
behavior of  the propagator of   two dimensional
free euclidean fermions, while
in Section \ref{ising} we study as
a further example of our approach
 the mass corrections to
the spin spin correlator in the Ising model.

\section{The general method and underlying hypotheses} \label{ipotesi}

We present in this Section the ideas of the approach
(see also \cite{zamo,sonoda1,mz}) and  state  a clear list of the underlying
hypotheses (which  are essentially three).

The goal is the reconstruction of the short distance behavior
of correlation functions
of composite operators of a $D$ dimensional euclidean  field theory
without dimensional couplings,
 perturbed by
one or more  operator  $\o_i$ (of
canonical dimension $x_i$)
  with  dimensionful
couplings $m^i$ ("generalized masses" of dimension $y_i\equiv D-x_i$):
\beq \Delta S= -\int\!\! dx \sum_i {m^i_B} {\o_i}_B(x), \eeq
(notice  the minus sign above, taken for
convenience).
We will refer to this model  as "deformed theory"
 while the unperturbed one will be called "massless theory".
The perturbation operator $\o_i$ is supposed here
relevant ($0<x_i<D$) to restrict the number
of renormalization conditions needed to define its  multiple insertions,
see Section \ref{uv}.

We will assume that all
 correlators of the deformed and of the massless theory  to
be well defined.

The massless theory is  supposed to be at least  perturbatively
U.V. renormalizable
with respect to some dimensionless coupling $\lambda$;
furthemore
we will require that only logarithmic corrections to tree level scale
invariance can arise.

The eventual
perturbative (with respect to $\lambda$) knowledge
of correlators at $m^i=0$   gives a perturbative estimation of
each term of the mass expansions of the correlators of the deformed theory;
nevertheless  in the particular case of (two dimensional)
conformal field theories
(bare) correlators are exactly known and thus
coefficients of mass expansion can be  {\it exactly} computed.
In the following we will forget the eventual dependence in $\lambda$
with the convention that all formulae should be intended to hold
order by order in renormalized perturbative expansion with respect to $\lambda$
in the general case.
Notice
that the absence of super-renormalizable couplings guarantees that
 I.R. problems do not arise in massless theory
if a renormalization scheme without additional unphysical masses is used
(see e.g. discussion in the second and third papers of \cite{apms}).

The short distance behavior of the deformed theory is described by
the  Operator
Product Expansion \cite{wilson}:
 \beq
 <[\Phi_{a_1}(r_1) \cdots \Phi_{a_n}(r_n)]  X>_m \sim
C_{a_1\cdots a_n}^c (r_1-r_n,\cdots r_{n-1} -r_n ;m)
<[\Phi_c (r_n)] X>_m,
\label{generalope}\eeq
where $\Phi_b$  are a complete set of
composite operators
 (that reduce in the limit $m^i \to 0$
to the  operators of the massless theory),
$X $ is a multilocal product of other fields
localized "far away"
from the $r_i$, the suffix $m$ refers to deformed theory,
and the convergence of the  series
will be discussed below. In the following we will denote the renormalized
multiple insertions of operators, see Section \ref{uv},
 enclosing them with $[\cdots ]$.

We will present  an I.R. finite technique
 to compute multiple derivatives of
Wilson coefficients $ C_{a_1\cdots a_n}^c$ with respect to
(renormalized) $m^i$ at the point
$m^i=0$ in terms of integrated correlators of the massless theory.
We will assume the following:

\smallskip
{\bf Hypothesis 1 (Regularity):} {\it an U.V.
renormalization scheme  for correlators
of the deformed theory is assumed
such that
counterterms are polynomial in
renormalized  generalized masses (Minimal Mass
Dependence, MMD).
Furthermore, the  Wilson coefficients are supposed to be regular
(i.e. ${\rm C}^\infty$)  in generalized masses
at $m^i=0$.
}

Minimal Mass Dependence  guarantees the smoothness of the
 $m^i\to 0$ limit from the point of view of
U.V. renormalization, property which will be essential in the following.
Any subtraction scheme that subtracts only the infinite
part in the underlying regularization (Minimal Subtraction) is expected to
satisfy MMD because no additional mass dependence
is introduced in counterterms (apart from the
trivial powerlike factors) that gives the correct physical dimension.

We remark that Wilson coefficients and Minimal Mass Dependence are not
 independent: at least
in the framework of perturbative renormalization of
quantum field theories MMD is equivalent  to regularity of
Wilson coefficients.
This can be understood as follows:
first we remind  that Minimal Subtraction in Dimensional
Renormalization (MS) satisfies MMD, \cite{thooft},
and regularity, \cite{ana}.
Secondly any  new scheme satisfying MMD
will
differ from MS by means of finite   counterterms, polynomial in the
masses.
In particular the change of scheme for renormalized operators
will amount to the transformation
\beq
[\Phi_a]' =N_a^b [\Phi_b]
\eeq
with $N_a^b=\delta_a^b + Y_a^b$ and $Y$ a matrix of order
$O(\hbar )$ that is polynomial in the masses.
Wilson coefficients will change accordingly as
\beq
{{C'}_{a_1\cdots a_n}^c}=N_{a_1}^{b_1} \cdots N_{a_n}^{b_n} {N^{-1}}_d^c
C_{b_1\cdots b_n}^d
\eeq
and the new ones will be clearly regular in $m=0$ as the ones in MS scheme.
Conversely in a scheme with non regular counterterms the matrix $N_a^b$
above will have some non regular  entry and so for Wilson coefficients.

We want to emphasize  furthermore  that the stronger
 (nonperturbative) assumption
of {\it analyticity} of Wilson coefficients in generalized masses
would ensure the {\it convergence } of the Taylor series
we are building up here, which is
only  asymptotic in the general case.

The second assumption necessary to compute derivatives with respect to
generalized masses is:

\smallskip
{\bf Hypothesis 2 (Action Principle):} {\it
 it is assumed that for each renormalized generalized mass
$m^i$ a  conjugate operator
$\o_i$  exists such that the derivative
with respect  to $m^i$ is realized as:
\beq\label{actionprinciple}
\de_{m^i} <[X]>_m = \int\!\! dx <[:\o_i(x): X]>_m \qquad
  :\o_i: \equiv \o_i- <[\o_i]>_m
\eeq
for each multilocal operator $X$.
}

The Action Principle \cite{schwinger}
 is well known to hold in perturbative renormalization,
 in the classical
B.P.H.Z. framework \cite{apbphz} as well as in other schemes
 satisfying
Hypothesis 1 such as  Dimensional Renormalization
\cite{apms} and Analytic Renormalization
 \cite{apan} (see also \cite{apdif} for Differential Renormalization).
 We will come back to this argument in Section \ref{uv}.

The simplicity of Eq.(\ref{actionprinciple}) compared to
"Variational
Formulae" \cite{sonoda1}
is the key of our approach that will allow us to deal with expressions for
derivatives of generic
order and to give the inductive IR finitess proof.

Let us finally have a glance at the method in a simple case, to introduce
 the final
hypothesis.
Suppose we want to compute the first derivative with respect  to $m^i$
(in $m^i=0$) of the two point
Wilson coefficient $C_{a b}^{1}$
(with $a,b,1$ referring to renormalized operators
$\Phi_a, \Phi_b, \1$).
Let us
 consider the quantity (remainder of O.P.E.)
\beq \label{deltadef}
\Delta^{(N)}(X_R,m)
\equiv <[(\Phi_a (r) \Phi_b (0) -
\sum_{c}^{x_c \leq N} C_{ab}^c \Phi_c(0))X_R]>_m
 \eeq
in which $x_c$ is  the
canonical dimension of operator $\Phi_c$ and $X_R$ is
the trivial operator ($X_R=\1$) or is a
multilocal operator with support in
 \beq
E_R \equiv \{ x / |x|>R \}.
\eeq
such that $R>|r|$.
(In the following we also
 denote by $I_R$ the complement set  of $E_R$, i.e. $I_R ={\bf R} ^D-E_R$).

We then take derivatives of both sides with respect  to $m^i$:
by using Action Principle we can write
\bea \de_i \Delta^{(N)}( X_R, m)&=&\int \!\!dx
< [:\o_i (x) : (\Phi_a \Phi_b -\sum_{c}^{x_c \leq N} C_{ab}^c \Phi_c )X_R]>_m
\non\\
 & &-\sum_{c}^{x_c \leq N} \de_i C_{ab}^c <[\Phi_c X_R]>_m
.\eea

It is evident that the in  the limit of zero generalized masses
the integral in the right hand side might be  I.R. divergent.
To deal with this problem
 we must proceed as follows, \cite{sonoda1}.
 First we split the integral in two pieces,
 \bea\label{twopieces}
 \de_i \Delta^{(N)}( X_R, m)&=&\int_{I_{R_1}} \!\! dx
< [:\o_i : (\Phi_a \Phi_b -\sum_{c}^{x_c \leq N} C_{ab}^c \Phi_c )X_R]>_m
\non\\
 &+&\int_{E_{R_1}} \!\! dx
< [:\o_i : (\Phi_a \Phi_b -\sum_{c}^{x_c \leq N} C_{ab}^c \Phi_c )X_R]>_m
\non\\
 & &-\sum_{c}^{x_c \leq N} \de_i C_{ab}^c <[\Phi_c X_R]>_m
,\eea
in which $R_1>R$.
Then we observe that the second integral can be rewritten as
\beq\label{badpiece}
\int_{E_{R_1}} \!\! dx
<[ :\o_i : (\Phi_a \Phi_b -\sum_{c}^{x_c \leq N} C_{ab}^c \Phi_c )X_R]>_m
 =  \Delta^{(N)}(  X_R', m)  \eeq
i.e. in the same form of Eq.(\ref{deltadef}) with
$X_R' \equiv\int_{E_{R_1}} \!\! :\o_i: X_R$ satisfying the same hypothesis
on the support as $X_R$.

If we could  assume
the (weak) convergence of  O.P.E. series when inserted
in correlation functions containing
"far" operators $X_R$ for the {\it deformed} theory, we would say
that the limit $N\to \infty$ of $\Delta^{(N)}( X_R, m)$
 of (\ref{deltadef}) is zero together
with all its mass derivatives and in particular
the $N\to \infty$ limit of  Eq.(\ref{badpiece}) would be zero for arbitrary
$R_1>R>|r|$: the subsequent limit
 $m\to 0$ of Eq.(\ref{twopieces}) would be safe
and could be exchanged with the first (I.R. cut off) integral with the
desired result
\bea
0&=& \lim_{N\to \infty} \{\int_{I_{R_1}} \!\! dx
< [:\o_i : (\Phi_a \Phi_b -\sum_{c}^{x_c \leq N}
 C_{ab}^c \Phi_c )X_R]>_{m=0}\non\\
& &-\sum_{c}^{x_c \leq N} \de_i C_{ab}^c <[\Phi_c X_R]>_{m=0}\}
\eea
(the suffix $<>_{m=0}$ will be omitted in the following).

We think that, in spite of some  argument in favor \cite{polyakov},
 convergence of O.P.E. is far from being reasonably proved in a
quantum field theory
with (generalized) mass; in particular the existence of terms of the
form $e^{-1/(mr^{y}) }$ could spoil the convergence of the
expansion for small $r$ (see the first paper in \cite{review2}).

Nevertheless for our purposes
it suffices the following weaker
 hypothesis that we state here for general Wilson
coefficients.

\smallskip
{\bf Hypothesis 3 (O.P.E. asymptotic weak convergence)}:
{\it it is assumed that the remainder of O.P.E.
\beq
\Delta^{(N)}(X_R,m) \equiv <[(\Phi_{a_1}(r_1) \cdots \Phi_{a_n}(r_n) -
\sum_{c}^{x_c\leq N} C_{a_1 \cdots a_n}^c  \Phi_c)X_R]>_m
\eeq
($X_R$ being the unity or an arbitrary multilocal operator with support
outside $R>|r_1|,\cdots |r_n|$)
satisfies
\beq\label{ope1}
\lim_{N\to \infty} \lim_{m \to 0} \lim_{R\to \infty}
\de_{i_1}\cdots \de_{i_k}\Delta^{(N)}(X_R,m)=0
\eeq
for each $k$. Equivalently
\footnote{
\label{smooth}
In the rest of the paper  we will exchange derivatives with respect
to generalized masses with the limit over I.R. cutoff when the limit
over $m$ follows, because at $m\neq 0 $ the limit over $R$ is smooth.
}
 we can write that
\beq\label{ope2}
\lim_{N\to \infty}\lim_{R\to \infty}<[(\Phi_{a_1}(r_1) \cdots \Phi_{a_n}(r_n) -
\sum_{c}^{x_c \leq N} C_{a_1 \cdots a_n}^c  \Phi_c)X_R]>\sim0
\eeq
in the sense of asymptotic series (in $m^i$).
}

First of all we notice that assumption Eq.(\ref{ope2}) for $k=0$ requires the
convergence of the O.P.E. in the underlying $m^i=0$ theory
when inserted in correlators with
"far" operators. This property is well
 known to hold in  conformal  field theories,
\cite{opeconf}.
Also from axiomatic \cite{ope} as well from perturbative \cite{perturbativeope}
considerations we can say that in general quantum field theories
 O.P.E. is an asymptotic expansion
in powers of  $r$. But in the limit $R\to \infty$, by clustering,
the contribution of $X_R$ in the correlator factorizes and
 the only physical scale in the O.P.E. becomes the generalized mass (the
renormalization point $\mu$ cancels between V.E.V. and Wilson coefficients
 by Renormalisation Group invariance of the sum)
and a remainder of order $O( r^\nu)$ is actually of the form
 $O( (m^{1\over y} r)^\nu)$
by dimensional analysis, and thus  the O.P.E series with respect to powers of
 the
generalized masses reasonably becomes  asymptotic.

Let us come back to our previous example.  In the limit of Eq.(\ref{ope1})
it is clear that the contribution from Eq.(\ref{badpiece}) vanishes.
Observing that also
$\de_i \Delta$ vanishes, and choosing $X_R=1$ we obtain the formula
\beq\label{quite}
0= \lim_{N,R_1\to\infty} \int_{I_{R_1}}
<[ :\o_i : (\Phi_a \Phi_b -\sum_{c}^{x_c \leq N} C_{ab}^c(0)\Phi_c )]>
-\sum_{c}^{x_c \leq N} \de_i C_{ab}^c(0) <[\Phi_c]>
\eeq
in which the limit over $m$ and $R$ were exchanged without
problems for the surviving I.R. regulated
quantities. (Notice also that $(0)$ in Wilson coefficients refers to $m=0$;
we will omit this  specification for simplicity of notations in next Sections,
being clear from the context if we
 are dealing with zero mass Wilson coefficients
 or not.)

To have a useful formula we must observe
that in general,
by dimensional considerations,
\beq\label{selectionrule}
\lim_{R_i\to\infty} \int_{I_{R_1}}  \!\! dx_1\cdots \int_{I_{R_k}}\!\!  dx_k
<[:\o_{i_1}: \cdots :\o_{i_k}: \Phi_c(0)]>=0
\eeq
if $ x_c- \sum_{j=1}^k y_j >0$ ($y_i$ being mass dimensions of $m^i$) due to
the
absence of a physical scale in the $m=0$ theory
(we are using the fact that renormalization point $\mu$
 does not give powerlike corrections).

By  Eq.(\ref{selectionrule}) the series
 in $c$ index is truncated and we can write
\beq\label{firstorder}
\de_i C_{ab}^1(0)= \lim_{R\to\infty} \int_{I_{R_1}} \!\! dx
<[ :\o_i : (\Phi_a \Phi_b -\sum_{c}^{x_c \leq  y_i} C_{ab}^c(0) \Phi_c )]>.
\eeq
The goal of expressing the (first) derivative of a Wilson coefficient in term
of
well defined and known quantities of massless theory
is then reached! In the next Section we
will deal with higher order derivatives.
Notice that all formulae that we will derive   {\it do not depend } on the
U.V. renormalization
scheme, provided that Hypotheses 1-3 are satisfied. The choice of the scheme
is a matter of taste and computational convenience.

\section{All order formulae and I.R. finitess proof} \label{ir}
In this Section we will give I.R. safe expressions for the  $n^{th}$ derivative
of the Wilson coefficients with respect to generalized
 masses by using the assumptions of previous Section.
Then some consequences of these identities will be discussed.

We will prove by induction on $n$
  the general form of $n^{th}$ derivative of (two
operators) Wilson
coefficient at zero masses and its I.R. finitess. Generalization to higher
operator Wilson coefficients is straightforward.

We want to prove that at order
 $n$ the following relation holds for the deformed theory:
\bea & &
 \lim_{N\to \infty}\lim_{R\to \infty}\de_{i_1}
 \cdots\de_{i_n} \Delta^{(N)}( X_R ,m )
\sim \non\\
& &\lim_{N,R\to \infty}\{
\int_{I_{1}}\!\! dx_{1}\cdots \int_{I_{n}}\!\! dx_{n}
 <[ :\o_{i_n}: \cdots:\o_{i_1}:
(\Phi_\au  \Phi_\ad  -\sum^{x_b\leq N}C_{\au \ad}^b \Phi_b) X_R]>_m
\non\\
& &
- \sum^{x_b\leq N}\de_{i_1} C_{\au \ad}^b \int_{I_{2}}\!\! dx_{2}\cdots
 \int_{I_{n}}
\!\! dx_{n}
 <[ :\o_{i_n}: \cdots :\o_{i_2}: \Phi_b X_R]>_m + {\rm p.} \non\\
& &
- \sum^{x_b\leq N}\de_{i_1} \de_{i_2} C_{\au \ad}^b
\int_{I_{3}}\!\! dx_{3}\cdots \int_{I_{n}}\!\! dx_{n}
 <[ :\o_{i_n}: \cdots :\o_{i_3}: \Phi_b X_R]>_m + {\rm p.} \non\\
& & \cdots
 - \sum^{x_b\leq N}\de_{i_1} \cdots \de_{i_n} C_{\au \ad}^b <[\Phi_b X_R]>_m
\}
\label{n=n}\eea
in which
$I_i\equiv I_{R_i}$ for shortness and $R_n,R_{n-1} \cdots R_1>R >|r|$.
The series
 are to be intended asymptotic as explained in the previous Section and with
{\it p.} we mean all terms obtained interchanging the
suffix $(1,\cdots ,k-1)$ of Wilson coefficients' derivatives with
$(k,\cdots ,n)$ inside integrals
without
distinguishing the ordering.
We observe that the left hand side of the previous expression is zero
asymptotically  due to Hypothesis 3,  Eq.(\ref{ope2}), and that due to the
presence of I.R. cutoff the massless limit is I.R. safe as we shall use
below.

It is clear by Hypothesis 3 that at $n=0$ Eq.(\ref{n=n}) holds.
 Now let us assume that
Eq.(\ref{n=n}) holds at order $n$ and show that it holds at order $n+1$.
Let us take derivative with respect
 to $m^{i_{n+1}}$  of both sides of Eq.(\ref{n=n})
(and commute derivative with
 respect to limit over $R$, see footnote \ref{smooth}).

The action of $\de_{i_{n+1}}$ on the correlators gives
terms of the form
\beq
\lim_{N,R\to \infty}\int \!\! dx_{{n+1}} <[:\o_{i_{n+1}}(x_{n+1}):
 \left\{ n \right\}]>_m
\eeq
i.e. the same terms of order $n$ (contracted with Wilson coefficients
as in (\ref{n=n})), indicated above with $\{n\}$,
  but with the additional insertion of $:\o_{i_{n+1}}:$, and terms in which
derivative acts on Wilson coefficients present at order $n$.
But  by  Eq.(\ref{n=n}) (with $X_R \to \int_{E_{n+1}}\!\!
 dx_{{n+1}}  :\o_{i_{n+1}}: X_R$)  we obtain
\beq
\lim_{N,R\to \infty}\int\!\! dx_{{n+1}} <[:\o_{i_{n+1}}: \left\{ n \right\}]>_m
=\lim_{N,R\to \infty}\int_{I_{n+1}}\!\! dx_{{n+1}}
<[:\o_{i_{n+1}}: \left\{ n \right\}]>_m
\eeq
if we restrict to $R_n ,R_{n-1},\cdots ,R_1\geq R\to \infty$.

It is easy to check that the sum of the two contributions simply reproduces
the $n+1$ expression according to (\ref{n=n}) so that induction works
{\it Q.E.D.}

If we choose $X_R=1$ in Eq.(\ref{n=n}), take the $m^i \to0$ limit
and  remind the dimensional selection rule Eq.(\ref{selectionrule}) we obtain
finally that, in the limit $R_n ,R_{n-1},\cdots ,R_1\geq R\to \infty$,
the series in index $b$ are actually finite sums and that
\bea
&& \de_{i_1} \cdots \de_{i_n}   C_{\au \ad}^1
=\non\\
& &\lim_{R\to \infty}\{
\int_{I_{1}} \!\! dx_{1}\cdots \int_{I_{n}}\!\!  dx_{n}
 <[ :\o_{i_n}: \cdots :\o_{i_1}:(\Phi_\au  \Phi_\ad
- \sum_b^* C_{\au \ad}^b \Phi_b) ]>
\non\\
& &
- \sum_b^*\de_{i_1} C_{\au \ad}^b \int_{I_{2}}\!\!  dx_{2}\cdots
\int_{I_{n}}\!\! dx_{n}
 <[ :\o_{i_n}: \cdots :\o_{i_2}: \Phi_b ]> + {\rm p.} \non\\
&&\cdots\non\\
& &
- \sum_b^* \de_{i_1} \cdots\de_{i_{n-1}} C_{\au \ad}^b
\int_{I_{n}}\!\! dx_{n}
 <[ :\o_{i_n}:  \Phi_b ]> + {\rm p.} \}
\label{main} \eea
where $\sum_b^* <[:\o_{i_n}:\cdots :\o_{i_k}: \Phi_b]>$
 is restricted to $\Phi_b$ such
$x_b\le y_{i_k}\cdots +y_{i_n}$.

More generally  if  we keep  in Eq.(\ref{n=n}) a nontrivial
 $X_R$
\footnote{
As an example, in conformal field theories the  particular choice
$X_R= \Phi_c(z,\bar z) z^\Delta {\bar z}^{\bar \Delta }$
(with $z,\bar z$ complex coordinates of modulus $R$
and $\Delta, \bar \Delta $ the relative
canonical
 dimensions of operator $\Phi_c$
in the underlying conformal field theory)
will select
$\de_{i_1} \cdots \de_{i_n}   C_{\au \ad}^c $ in
 the last term of  Eq.(\ref{mainmain}), by  orthonormality
property of conformal operators.}
and take the $m^i\to 0$ limit,
we can observe
that
\beq
\int_{I_{k}}\!\!\cdots\int_{I_{n}}\!\!
 <[:\o_{i_n}:\cdots :\o_{i_k}: \Phi_b(0) X_R]>
=O(R^{\delta})
\eeq
in which we defined the I.R. degree of divergence of the correlator as
\beq\label{irdeg}
\delta\equiv\sum_{j=k}^n y_{i_{j}}-x_b-x_{X_R}
\eeq
by simple dimensional considerations.
It  follows that all correlators
 with $\delta<0$ give vanishing contribution in the
$R\to \infty $ limit and the series in the index $b$ are effectively truncated.

We obtain in all generality:
\bea
0&=&
\lim_{R\to \infty}\{
\int_{I_{1}}\!\! dx_{1}\cdots \int_{I_{n}}\!\! dx_{n}
 <[ :\o_{i_n}: \cdots :\o_{i_1}:(\Phi_\au  \Phi_\ad
- \sum_b^* C_{\au \ad}^b \Phi_b) X_R]>
\non\\
& &
- \sum_b^*\de_{i_1} C_{\au \ad}^b \int_{I_{2}}\!\! dx_{2}\cdots
\int_{I_{n}}\!\! dx_{n}
 < [:\o_{i_n}: \cdots :\o_{i_2}: \Phi_b  X_R]> + {\rm p.} \non\\
&&\cdots\non\\
& &
- \sum_b^* \de_{i_1} \cdots\de_{i_{n-1}} C_{\au \ad}^b
\int_{I_{n}}\!\! dx_{n}
 <[ :\o_{i_n}:  \Phi_b X_R ]> + {\rm p.} \non\\
& & -\sum_b^* \de_{i_1} \cdots \de_{i_n}   C_{\au \ad}^b <[\Phi_b X_R]> \}
\label{mainmain} \eea
where $\sum_b^* <[:\o_{i_n}:\cdots :\o_{i_k}: \Phi_b X_R]>$
 is restricted to $\Phi_b$ such
$x_b\le y_{i_k}\cdots +y_{i_n}-x_{X_R}$.

Eq.(\ref{mainmain}) gives  an infinite number of
linear algebraic relations (one for each choice
of $X_R$) among derivatives of Wilson coefficients up to order $n$.
Of course consistency of
the method (guaranteed by our assumptions) enforces that
only a finite number of relations are independent and the system is not
overconstrained.

Moreover it is important to observe that in {\it each}
 equation such as (\ref{mainmain}) the I.R. divergences of the same form cancel
{\it independently}.
 In particular  this happens for terms of the same
dimensionality (I.R. degree of divergence): if for fixed $a_1, a_2$
we define the finite set
\beq
S_\sigma \equiv \{ \de_{i_1}\cdots \de_{i_k} C_{a_1,a_2}^b /
\sigma=\sum_{j=1}^k y_{i_{j}}+x_b \}
\eeq
it is clear that for every choice of $i_1, \cdots, i_n$ and $X_R$
in Eq.(\ref{mainmain}) we have
linear relations only  between elements
of the same  sets $S_\sigma$, because
$\delta=\sum_{j=1}^n y_{i_{j}}-x_{X_R}-\sigma$
for each term of (\ref{mainmain}).

At this point to compute a derivative of a Wilson coefficient belonging to
a set $S_\sigma$ one has only to choose a convenient number of subrelations
of (\ref{mainmain}) (varying $X_R$ or $n$) and solve a linear system.
Notice that due to possible different form of I.R. divergences with the
same dimensionality the number of relations obtained by I.R. cancellation
might become even bigger. This can happen also when some quantum numbers are
conserved.

We want to show  now that, while  the particular I.R. spatial cutoff we used
was essential to derive Eq.(\ref{n=n}) by induction, our final results,
Eq.(\ref{main}) and Eq.(\ref{mainmain}) hold for a wide class of I.R.
regularizations.
The point is that in Eq.(\ref{n=n}) all the terms have the form (keeping
only essential features)
\beq
 \lim_{m\to 0} \lim_{R_i\ge R\to \infty}
 \int \!\! dx_1 \cdots \int\!\!  dx_k
\theta_{R_1} (x_1) \cdots \theta_{R_k}(x_k)
<\cdots X_R>
\eeq
where $\theta_R(x)=\theta(|x|-R)$ (with $\theta$ the usual step function).
Let us now  consider a generic I.R. regulator function $\Theta_{R'}(x)$
($R'$ being a new length parameter)
such that
$\lim_{R'\to \infty}\Theta_{R'}(x)=1$. Inserting $1$ in the integrals and
 exchanging
integration with these limits we have
\beq
\lim_{m\to 0} \lim_{R_i\ge R\to \infty}
\lim_{R_i'\to \infty}
 \int \!\! dx_1 \cdots \int\!\!
   dx_k \theta_{R_1} (x_1) \cdots \theta_{R_k}(x_k)
\Theta_{R_1'} (x_1) \cdots \Theta_{R_k'}(x_k)
<\cdots X_R>
.\eeq
It is clear that at $m\ne 0$ (the limit on $m$ being external) we can
exchange limits over $R,R_i$ with those over $R'$ and subsequently
the limit over $R$ with integrations, obtaining
\beq
\lim_{m\to 0}
\lim_{R_i'\to \infty}
 \int \!\! dx_1 \cdots \int dx_k
\Theta_{R_1'} (x_1) \cdots \Theta_{R_k'}(x_k)
<\cdots X_\infty>
.\eeq

Proceeding exactly as done previously (exchanging limit over masses
 and over $R'$,
using
selection rules at $m^i=0$)
we obtain for a generic I.R. cutoff function $\Theta_R$:
\bea
0&=&\lim_{R_i\to \infty}\{
\int \!\! dx_{1}\cdots \int \!\!
 dx_{n} \Theta_{R_{1}} (x_{1})
 \cdots \Theta_{R_{n}}(x_{n})\times  \non\\
& &
 <[ :\o_{i_n}: \cdots :\o_{i_1}:(\Phi_\au  \Phi_\ad
- \sum_b^* C_{\au \ad}^b \Phi_b) X_\infty]>
\non\\
&-&
\sum_b^*\de_{i_1} C_{\au \ad}^b
\int\!\! dx_{2}\cdots \int \!\!
 dx_{n} \Theta_{R_{2}} (x_{2})
 \cdots \Theta_{R_{n}}(x_{n})
  \times\non\\
& &< [:\o_{i_n}: \cdots :\o_{i_2}: \Phi_b  X_\infty ]>
 + {\rm p.} +\cdots\non\\
&- &
 \sum_b^* \de_{i_1} \cdots\de_{i_{n-1}} C_{\au \ad}^b
\int \!\!
 dx_{n}
 \Theta_{R_{n}}(x_{n})
 <[ :\o_{i_n}:  \Phi_b X_\infty ]> + {\rm p.} \non\\
& -& \sum_b^* \de_{i_1} \cdots \de_{i_n}   C_{\au \ad}^b <[\Phi_b X_\infty ]>
\}
\label{mainmainmain} \eea
(same meaning of the $^*$ as before).
It is clear that a clever choice of I.R. cutoff can considerably
simplify computations. For instance it is  easy to prove that when
 $X_\infty=1$ and the cutoff
function $\Theta_R$ is rotation
invariant, only scalar operators are involved in the expressions.
This property will be used in Section \ref{example}

We close the Section noting that, while Eqs.(\ref{main}), (\ref{mainmain}),
 were derived by using only
of the condition
$R_n ,R_{n-1},\cdots ,R_1\geq R\to \infty$, with the result of having symmetric
formulae,  for practical calculations it is better
\cite{sonoda1} to force the I.R. cutoff
to go to infinity in a hierarchical way, say,
 $R_n >>R_{n-1},\cdots >>R_1\geq R\to \infty$
so that we can perform first the integral with
bigger cutoff $R_n$, neglecting subleading
 terms of order $O({R_{n-1} \over R_n})$
and so on.
This could not be the most convenient way to compute integrals in other I.R.
regularizations:
for instance in the case $\Theta_{Q_i}(x)= e^{iQ_ix}$ it is simpler to take
all
$Q_i=Q\to 0$.

\section{U.V. definition of composite operators}\label{uv}
In this Section we schematically review some known facts on the treatment of
renormalized multiple insertions of
composite operators with particular
 reference to the operator $\o_m$ conjugate to
the generalized mass $m$ by the Action Principle Eq.(\ref{actionprinciple})
(we will deal here only with one generalized mass; extension to general
case should be straightforward for the reader).

The foundations of  renormalization are locality and power counting
(\cite{bogo}, see \cite{becchi} for a pedagogical introduction).
Each  composite operator $\Phi_a$ has a related power counting dimension
 equal  to
the canonical one $x_a$, (we will  not consider oversubtracted operators here).
 Each insertion of $\Phi_a$ in a truncated connected  one particle
irreducible graph $\gamma$ changes the U.V. superficial
degree  of divergence $\omega_\gamma$
 by the amount
$\delta \omega=-y_a=-D+x_a$,
generating new divergences.
Divergences related to one operator insertion are treated
allowing for a
mixing of each  bare operator with those of lower power counting dimension
to get a finite renormalized one,  $[\Phi_a]$.
Renormalized multiple insertions of composite operators
(denoted here by $[\cdots ]$) contain
new divergences that should be cured by counterterms proportional to delta
function and its derivatives, the so called contact terms: for instance
\beq
[\Phi_a(x) \Phi_b(0)]=[\Phi_a(x)][\Phi_b(0)]+{\rm contact \; terms}
.\eeq
For detailed treatment of arbitrary multiple insertions  see \cite{collecott}
and references therein.
A practical way to deal with multiple insertions
is obtained introducing sources $\omega^a(x)$ of dimension $y_a$
for composite operators $[\Phi_a]$, and allowing for counterterms
proportional to monomial in the sources, in the composite fields and
  their derivatives
of total dimension less then or equal  $D$ (see e.g. \cite{zinn,shore}).

For a composite operator to be well
defined a { normalization condition} should be given in correspondence of
each infinite subtraction.
A {\it renormalization scheme}
is a consistent and physical
  way to subtract divergences
and to fix finite counterterms with explicit or implicit normalization
conditions.

Equivalently,
a renormalization scheme can be seen  as
a  consistent technique to build up well defined tempered distributions
(satisfying physical requirements, \cite{bogo}) from "bare expressions" that
do not define a distribution and to fix
 finite counterterms arising from
the ambiguities of this
extension. This point of view is more suitable for
conformal field theory where "bare" exact expressions
are known for correlators of composite operators at different positions.

For instance the bare correlator with two insertions of the operator
$\e\equiv -:\bar{\psi} \psi:$
in the free massless $D=2$ euclidean Dirac
Fermions model (that is a conformal field theory)
is
\beq
< \e(r) \e(0)>_B= {1\over 2 \pi^2} {1\over |r|^2}
\eeq
but $1/r^2$ {\it is not } a distribution.
This can be understood because it
is not locally integrable with respect to the measure $d^2r$.
A distribution $T(r)$ that extends the bare correlator can be defined
(in polar coordinates)
as the solution of the distributional equation
\beq
|r| T(r)={1\over 2\pi |r|}
\eeq
($1/r$ {\it is } a distribution) whose general solution will have the form
\beq
T(|r|)=T_P(|r|)+ T_H(|r|)
\eeq
where $T_P$ is a particular solution of the complete equation such as
$Fp{1\over 2\pi |r|^2}$ (Hadamard finite part, see \ref{finitepart})
and $T_H=C_0 \delta(|r|)$ is
a solution of the homogeneous equation. The renormalized correlators
will be equal to $T$ and the dimensionless finite contact term $C_0$ will
be fixed after imposing
a normalization condition. The alternative definition
\beq
|r|^n T(r)={1\over 2\pi } |r|^{n-1}
\eeq
for some $n>1$ would give a $T_H=\sum_{k=0}^{n-1} C_k\delta^{(k)}(|r|)$ and
would change drastically
the short distance behavior of the correlator: in this sense is not
 physically relevant.

{}From our  discussion  it appears clearly
that only a finite number of insertions
of a relevant operator give U.V. divergences
 (and contact terms  to be eventually fixed): this is the main practical
motivation to restrict to relevant perturbations, while  all
formulae presented in Section \ref{ir}
can be extended formally to any marginal operator
(at any finite order).

After this brief digression let us come to the problem of defining
renormalized operators (i.e. of fixing schemes)
satisfying the
Action Principle.
The point is that, while being clearly true at bare level
(think  about functional integration), the Action
Principle might be not satisfied in a particular scheme,
being possibly violated by renormalization. In this case
the starting point scheme must be corrected  (i.e. changed) by finite
 counterterms
to enforce AP.
It follows in our case that  all the U.V. divergent
insertions of the renormalized composite operator $\o_m$
to be cured in a chosen scheme
might require additional contact term to be fixed in such a way
that  Eq.(\ref{actionprinciple}) holds. These  contact terms
result to be
finite
numbers (apart the trivial power of $m$) whose expression in
terms  of integrals of
U.V. renormalized correlators (in the original scheme)  is known
and the only free parameter
 characterizing
the composite operators is the usual renormalization point $\mu$.

Of course  it is possible that in some renormalization
 scheme the Action Principle
holds {\it automatically} without finite renormalization;
this is the case of Dimensional Renormalization,
\cite{apms}
 where no additional contact terms
 are needed and life is easier.

In our approach we are interested only in the zero mass limit,
thus all dimensionful contact
 terms vanish and are not involved in our expressions.
Only dimensionless  contact
  terms could survive  in zero mass correlators: they
have a {\it definite} expression in terms of massive integrals but cannot
be computed in terms of the massless quantities only. This situation arises
when one tries to impose the  condition
\beq\label{recondi}
\de^k_m  <[\Phi_a]>_m=
\int\!\! dx_1 \cdots dx_k <[\o_m(x_1)\cdots \o_m(x_k) \Phi_a]>_m
\eeq
in the case
 when $x_a=k y_m$ (in general when  the I.R. degree of divergence $\delta$
defined in Eq.(\ref{irdeg}) is zero).
Equation (\ref{recondi}) presents non trivial U.V. renormalization
and also  is I.R. singular at
$m=0$ ($\log m$ terms are clearly present in the operator V.E.V.):
it follows that
   a numerical constant that parameterizes
the I.R. behavior of the operator $\Phi_a$ in the deformed theory
is left not fixed
(but expressed in terms of deformed theory correlators).

These unknown constants
will parameterize our results.
 In the case of one generalized mass we can always
confine this ambiguity to the composite operators V.E.V., as will be shown in
the next Section with two examples. See also the related paper \cite{gth}

Before closing this Section
 we comment here that
"Variational Formulae"
\footnote{
In our mind "Variational Formulae"
are incomplete because do not give a general definite prescription to build up
 distributions from bare correlators (i.e. do not define a scheme).
 In particular it seems to us that
there is no prescription  to subtract singularities arising when three or more
operators collapse to a point, as happens at high derivative orders.}
of Ref. \cite{sonoda1} were invented
with the goal of building up a  geometrization
of the theory space of the renormalized theories, \cite{sonoda3}, with
a connection expressed
in terms of a set of   finite arbitrary  "contact terms".
We feel that the idea of geometrization (interesting by itself)
 is somewhat incompatible with
straightforward calculations and in our pragmatic approach we will
try to minimize the number of contact terms.
In particular a locally
 flat coordinate system (no connection at all) is obtained
when schemes
with a built in Action Principle  can be used.

\section{Two examples}
\label{example}

\subsection{A toy model example: short distance behavior of two dimensional
free fermion
propagator}\label{toy}
In this subsection we compute the mass corrections
to the short distance behavior of the two dimensional
euclidean free Dirac fermions propagator
up to order $m^2$ by using  the knowledge of the massless theory. We will
use as renormalization scheme the Minimal Subtraction
(MS) in Dimensional Regolarization.

To fix notations (we use that of \cite{zinn})
we write here the exact expression for the propagator we want to recover:
\beq
<\psi(r) \bar{\psi}(0)>_m=-{m\over 2\pi} K_0(m|r|) -{m\over 2\pi}
{{\hat r}\over |r|} K_1(m|r|)=\int\!\!
 {d^2p \over (2\pi)^2} e^{-ipx} {1\over i \hat p -m}
\eeq
in which ${\hat r}=r^\mu \sigma^\mu$, $\mu=1,2$.

We want to use Eq.(\ref{main}) with $n=1,2$,  that we rewrite here
 (neglecting
the trivially zero contributions and changing slightly notations):
\bea
& &\de_m C^1(r)=\lim_{R_1\to \infty} \{ \int^{R_1}\!\! d^2x_1
<[:\o_m(x_1): ( \psi(r) {\bar \psi} (0) - C^{\o_m} \o_m)]>\}\label{first}\\
& &
\de_m^2 C^1(r)=\lim_{R_2>>R_1\to \infty}
\{ \int^{R_1}\!\! d^2x_1\int^{R_2} \!\! d^2x_2 <[:\o_m(x_1)::\o_m(x_2):
(\psi(r) {\bar \psi} (0)\non\\
& &- C^1- C^{\Phi_2^c} \Phi_2^c)]>
-\int^{R_1} \!\! d^2x_1 \de_mC^{\o_m}  <[:\o_m(x_1)::\o_m:]> -
 (R_2 \leftrightarrow R_1)
\non\\
\label{second}
 \eea
where $\o_m$ and $\Phi_2^{c}\equiv (\bar\psi\Gamma^c\psi )
 (\bar\psi\Gamma^c\psi )$
are the marginal scalar operators of the theory
($\Gamma^c=(1,\gamma_\mu, \sigma_3)$).
Notice that lower indices of Wilson coefficients have not been reported.

We consider this example very instructive because the values of
the (differentiated)
 zero mass Wilson coefficients $ C^{\o_m}$,$C^{\Phi_2^c}$,$\de_mC^{\o_m} $
will be fixed imposing cancellation of I.R. divergences (of course the
first two can be obtained in independent way from the massless theory).

In principle we  should give proper normalization
conditions for
the one and two insertions
of the operator $\o_m\equiv\bar{\psi} \psi$ in such a way that
the Action Principle (Eq.(\ref{recondi})
with $k=1$
and $\Phi_a=\o_m$) holds.
The point is that  Action Principle is automatically satisfied
in the massive theory if the
 usual MS definition for composite operators is adopted
(see e.g. \cite{collins}).
If we use for a moment  the knowledge of the
massive theory propagator, we obtain after
 standard calculations:
\bea
& &<[:\o_m(r): :\o_m(0):]>_m = \int \!\! {d^2p \over (2\pi)^2}  e^{-ipr} F(p,m)
\label{oo}\\
& & F(p,m)\equiv -{1\over 2\pi}\int_0^1 \!\! dx
\left(\gamma_E+3+\log ({m^2+x(1-x)p^2\over 4\pi\mu^2})
\right) \non\\
& & <[\o_m]>_m
=-{m\over 2\pi}(\gamma_E+1+ \log ({m^2\over 4\pi\mu^2}) )\label{vev}
\eea
It is  easy to check that
\beq
\de_m<[\o_m]>_m = \int\!\! d^2r <[:\o_m(r): :\o_m(0):]>_m=F(0,m),
\eeq
i.e. Action Principle holds in this case without addition of contact terms.

Having we used only the knowledge of the massless theory, we would get
Eq.(\ref{oo}) with
\beq
F(p)\equiv F(p,0)= -{1\over 2\pi}(\gamma_E+1+ \log ({p^2\over 4\pi\mu^2}) )
.\eeq
We could also get a partial information on $<[\o_m ]>_m$ imposing that
the $\mu$ dependence
 (i.e. the U.V. behavior or in
other words the anomalous dimensions) is
the same in massive and massless theory:
\beq \label{diff}
\mu \de_\mu \de_m <[\o_m]>_m =
\mu \de_\mu \int^{R}\!\! d^2x <[:\o_m(x): :\o_m(0):]>
\eeq
(notice the $m$ dependence of  left side).
The cutoff integral can be computed exactly by using the
 properties of Bessel functions
\cite{gr}:
\bea & &
 \int^R \!\! d^2x <[:\o_m(x): :\o_m(0):]>
=\int\!\! {d^2p \over (2\pi)^2} F(p)
{2 \pi R\over p} J_1(pR) \non\\
& &={1\over 2\pi} (\log (\pi \mu^2R^2 )+\gamma_E -1) ) \label{bessel}
\eea
Integrating the differential equation (\ref{diff})
 and imposing the dependence on $m/\mu$ we get
\beq
\de_m<[\o_m]>_m = - {1\over \pi} \log{ m\over \mu}+ {\rm const},
\eeq
from which, integrating again, we get
\beq\label{omvev}
<[\o_m]>_m =- {1\over \pi} m (\log{ m\over \mu} +{\cal C}_{\o_m }).
\eeq
The constant ${\cal C}_{\o_m }$ can be fixed only imposing the Action
Principle
on the deformed theory (this gives the correct V.E.V.,  Eq. (\ref{vev}))
and its existence  has been anticipated
in Section \ref{uv}
(notice that the two point correlator has zero
I.R. degree of divergence and we expected I.R. problems
in this case).

We  computed some of the required integrals  passing to complex
coordinates $r=r^1+ir^2 \cdots$
and using Stokes' Theorem to reduce surface integrals to line integrals.
In  \ref{stokes} we give an explicit example of such calculations .

The results (up to not relevant
  subleadings terms in the limit $R_2 >>R_1>>|r|$)
 are:
\bea \int^{R_1} \!\! d^2x_1
<[:\o_m(x_1):  \psi(r) {\bar \psi} (0)]>
&\sim& -{1\over 2\pi}\log {R_1\over |r|}\non\\
 \int^{R_1}\!\!  \int^{R_2}\!\! <[:\o_m(x_1)::\o_m(x_2):
(\psi(r) {\bar \psi} (0)-C^1(r))]>
&\sim&
\non\\
 -{{{\hat r}}\over 8\pi}\{ \log{|r|^2\over R_1^2} +\log{|r|^2\over R_2^2} -2\}
  & &         \label{results2}
\eea

For what concerns the remaining integrals, it suffices to note that
\beq
\int^{R_1}\!\!\int^{R_2}\!\!<[:\o_m:(x_1):\o_m:(x_2) \Phi_2^a(0)]>
\propto  \log \mu R_1 \log \mu R_2 + \cdots
\label{results1}
\eeq
as it is easy to check.

Equating terms of same behavior in $R$, by  Eq.(\ref{bessel}),
Eq.(\ref{results1}), Eq.(\ref{results2}),
we get from Eq.(\ref{first})
\beq
C^{\o_m}=-{1/2} \;\;\; \de_mC^1 ={1\over 4\pi}\left(\log(\pi\mu^2 |r|^2)
 +\gamma_E -1\right)
\eeq
and from Eq.(\ref{second})
\bea
& & \de_m C^{\o_m} ={1\over 4}\hat r \qquad C^{\Phi_2^a}=0 \non\\
& &\de_m^2 C^1 =-{1\over 4\pi}
\left( \log \pi \mu^2 |r|^2 -2+\gamma_E \right) {\hat r} .
\eea
Combining these results  with the VEV of $\o_m$  Eq.(\ref{omvev})
(and with the off critical information (\ref{vev})), we get
\bea\non
<\psi(r) \bar{\psi}(0)>_m &\sim& -{1\over 2\pi} {{\hat r}\over |r|^2}
\left(1+{1\over 2}m^2|r|^2
 (\log ({m|r|e^{\gamma_E}\over 2})-{1\over 2})\right)\\
& &
+{m\over 2\pi}\log ({m|r|e^{\gamma_E}\over 2}) +O(m^3)
\eea
that is the correct result as can be
 checked from the properties of $K$ functions.
Another check  can be obtained from the constraints
\bea
& &\mu{d \over d\mu} \de_m^n C^1+n \de_m^{n-1} C^{\o_m}
\mu{d\over d\mu}<[\o_m]>_m =0
\;\non\\
& &\mu{d \over d\mu} \de_m^n C^{\o_m}=0
\eea
obtained from the free theory relation
\beq
0=\mu{d \over d\mu}
 <\psi \bar\psi>_m=\mu{d \over d\mu} (C^1(m) +C^{\o_m}(m)<\o_m>_m)
\eeq
equating terms of same order in $m$ and $\log m$.
\subsection{Application to Ising Model}\label{ising}
We want to reconstruct here the mass corrections to short distance behavior
 of the two points spin-spin correlation function of the Ising Model
by using   the previously  introduced method to compute Wilson coefficients.
Similar results have been already obtained in a previous article \cite{mz};
 we
only want to treat this model as a concrete example of our reformulation
to be compared with the preexisting approach.
To put in evidence the renormalization scheme
 independence of our treatment (provided Hypotheses 1-3
are satisfied), we will use here the Hadamard finite part (see
\ref{finitepart})
to deal with U.V. divergence and the  I.R. regulator $\Theta(x)=J_0(|Q| |x|)$
(see Eq.(\ref{mainmainmain})) that is equivalent to insert $\Theta(x)=e^{iQx}$
and then average over directions
 of $Q_\mu$ before taking the limit $Q\to 0$.

It is a well known fact \cite{bpz} that the Ising model in  proximity of
 its critical point
is described by  a theory of free Majorana fermions with mass $m$ and
euclidean action
\beq
S={1\over 2\pi}\int\!\! d^2x
 \psi {\bar \de} \psi + {\bar \psi} \de  {\bar \psi}+
im {\bar \psi}\psi.
\eeq

Moreover the exact correlator \cite{wu} is known
from the scaling limit of the lattice theory
and can be used as a check.

The  conformal theory at fixed point $m=0$
is  described by the primary operators
$\1,\s , \e \equiv : {i\over 2 \pi} {\bar \psi} \psi : $
 of
 dimension $x=0,1/8,1$ from which the secondary operators
can be built up by repeated application of the Virasoro generators:
\beq
\Phi^{ \{ n \} \{  {\bar n } \} }
= L_{-n_1} \cdots L_{-n_N}
 {\bar L}_{-{\bar n}_1}\cdots {\bar L}_{-{\bar n}_M} \Phi,
\eeq
where $\Phi$ is any of the primary operators and the
dimension of the secondary operator is $x=x_\phi+\sum n_i +\sum {\bar n}_i$.
Away from the fixed point the expression of the operator conjugate to $\de_m$
 will be
of the form $[\om ]= -\e + <[\o_m ]>1$.

At first order  we have
\beq
\label{goal}
\dm C_{\s \s}^1(r)=\lim_{Q\to 0}
 \int\!\! d^2x e^{iQx}<[ -\e (x) (\s (r) \s(0) -C_{\s \s}^{\e}(r) \e (0))]>
\eeq
(the average over directions
of  $Q_\mu$ to be taken after integrations is not reported for simplicity
of notations).

Before proceeding to the computation of the integrals a precise
definition
of the operator $\om$ should be given in order to obtain
well defined distributions from its insertions,
as discussed in Section \ref{uv}.
We define the double insertion
of $\om$ by
\beq \label{fp}
<[\om (r) \om(0)]>_m = \fp {m^2 \over 4 \pi^2} (K_1^2(mr) -K_0^2 (mr) )
\eeq
 where $\fp$ means the distribution obtained  taking the Hadamard finite
part of the U.V. singularity. (We used some additional
knowledge of the massive theory: the massless correlator can be derived by
taking the limit $m\to 0$ of previous expression).

Proceeding  as in Section \ref{uv} and imposing the Action Principle
\beq \label{results}
\dm <[\om ]>_m= \int\!\! d^2r <[\om (r) \om (0)]>_m=
-{1 \over 2 \pi}  (\log ({m\over 2\mu}) +1 -\psi(1))
\eeq
(see (\ref{massiveint})
for computation of the integral),
  we derive the V.E.V.:
\beq
<[\om ]>_m=-{m\over 2\pi} \log ({m e^{\gamma_E}\over 2 \mu})
.\eeq
We recall
 that
 (as explained in Section \ref{uv} and in Section \ref{toy}) the
 numerical constant inside the
logarithmic term cannot be fixed only by knowledge of the $m=0$ theory.

We can now come back to the computation of integrals in Eq.(\ref{goal}).

The (locally integrable)
correlation functions involved in the first integral is exactly known
from the conformal theory \cite{zuber}:
\beq
<\e(r_1) \s(r) \s(0) >= {r^{3/4}\over 4\pi |r_1-r| |r_1|} \label{exact}
\eeq
and the integral can be  computed (in the limit $Q \to 0$ )
 by standard Feynman diagram techniques and properties of Bessel Functions
\cite{gr}:
\bea
& &\int\!\! d^2x{1\over |x-r|}{1\over|r|}e^{iQx}=
\int_0^1\!\! dt{e^{i (Qr) t}\over (t(1-t))^{1/2}}
\int_0^\infty
\!\! {ds\over s}e^{-s} e^{-{t(1-t)Q^2 r^2/(4s)}}\non\\
& &=2\int_0^1\!\! dt{e^{i (Qr) t}
\over (t(1-t))^{1/2}} K_0({\sqrt {t(1-t)}} |Q| |r|)
\sim -2\pi \log{|Q||r|e^{\gamma_E}\over 8}
\label{i1}
\eea

For what concerns the second integral in (\ref{goal})
we have (after the limit $m\to 0$ is taken)
\bea
\int \!\!d^2x e^{iQx} < [\e (x)\e (0)]>
&=& \int\!\! d^2x  e^{iQx}\fp {1\over 4 \pi ^2 x^2 }
\non\\ &=&
{-1\over  2\pi }(\log {|Q|\over 2\mu}+\gamma_E)
\label{i2}\eea
(see \cite{schwartz} for the Fourier transform of Hadamard finite part).
Substituting (\ref{i1})-(\ref{i2}) in Eq.(\ref{goal}) we get
\beq
\dm  C_{\s \s}^1 = {1\over 2} r^{3/4} (\log{|Q||r|\over 2} +\gamma_E-\log 4)
-C_{\s \s}^{\e} {1\over  2\pi } (\log {|Q|\over 2\mu}+\gamma_E)
.\eeq
Cancellation of the $\log Q$  factor
fixes
\beq\label{ce}
C_{\s \s}^{\e}(r)=\pi r^{3/4}
\eeq
while equating the regular term we get
\beq\label{dc1}
\dm  C_{\s \s}^1 = r^{3/4} {1\over 2} \log ({\mu r\over 4} )
\eeq

{}From  (\ref{dc1}) we can thus reconstruct at first order in $m$
the deviation from the scaling behavior of the fixed point theory
for the  Wilson coefficient
\beq
C_{\s \s}^1 (r,m;\mu)={1\over r^{1/4}}
(1 +{mr \over 2} \log({\mu r\over 4} ) + O((mr)^2) )
\eeq
and (using also (\ref{ce})) of the short distance behavior of the
two point function:
\bea
<\s (r) \s (0) >_m &\sim& C_{\s \s}^1 (r,m;\mu)
+m \dm C_{\s \s }^m(r,m=0) <\om >_m\non\\
&=&
 {1\over r^{1/4}}(1 +{mr \over 2} \log({m r\over 8} e^{\gamma_E})+ O((mr)^2)).
\eea
These results are in agreement with those of \cite{mz}
 and  with the exact
ones
of \cite{wu}.

\section{Conclusions}  \label{conclusions}
We presented a consistent method
to deal with
I.R. singularities of expansions
in a relevant parameter around a field theory without
super-renormalized couplings.
Our approach is a reformulation
and an extension of other recent works
\cite{zamo,sonoda1,mz}.
The goal is the reconstruction of  the short distance behavior
(Wilson coefficients) of
the deformed theory
 obtained by adding a
 relevant operator to the action.
The  use of Action Principle
and (weak) convergence property of O.P.E.
give closed and I.R. finite expressions
 for the derivatives with respect to
generalized masses of  Wilson coefficients that,
 can thus be reconstructed by means of
Taylor expansion (that is convergent if analyticity of Wilson
coefficients can be assumed, otherwise asymptotic).

Our  contribution is an orthodox treatment of U.V. divergences of
composite operators that, as
a consequence, allows for an explicit and simple
 inductive proof
of  I.R. finitess  and  of general equation  Eq.(\ref{mainmainmain}).
 We put in evidence that  Eq.(\ref{mainmainmain}) does not refer
 to a particular U.V. renormalization scheme
(provided Hyp.1-3 are satisfied) and holds for general I.R.
regulators.

We discussed also the  presence of well defined but unknown constants
related  to validity of
renormalized Action Principle
that parameterize the long distance  behavior of deformed theory
(essentially operators' V.E.V.) and
cannot be computed by using the knowledge of the massless theory only.
Nevertheless these constant are
{\it universally} present in
 short distance expansions  of different products of composite operators
and are always in finite number
at any finite order in the masses.
The possibility of fixing these parameters (e.g. by
some symmetry relation between operators in the deformed theory or the
knowledge of
some particular deformed correlator, as in the Ising model of Section
\ref{ising})
depends on particular  models and should  be discussed case by
case.

In spite of this problem, we
think
that its simplicity and generality,
compared to many specific approaches,  make of this method a powerful
and promising tool.

We remind in particular
that the (three) main hypotheses on which it relies have been
 found to be  satisfied for the bulk of the interesting
conformal field theories if adequate renormalization schemes are used:
it thus seem that the technique could be helpful in the investigation
of relevant deformations of those theories.
 We hope to come back to the problem soon.

\smallskip {\bf Acknowledgements:} The authors are grateful to  C.M. Becchi
 for suggesting to them the problem and for many useful
discussions.
 We acknowledge also
for interesting discussions
 R. Collina, G. Delfino,
K. Konishi, C. Imbimbo,
Hiroshi Suzuki,
J.-B. Zuber. Part of the work  of one  author (R.G.)  was realized
during visits to S.Ph.T. of Saclay
and C.E.R.N., whose staffs are acknowledged for the warm hospitality.

\appendix

\section{Complex integrals}\label{stokes}
We give here
  an example of the computations required to obtain Eqs.(\ref{results})
the detailed calculation of the integral
\beq
I\equiv \int_{|z|\leq R} \!\! d^2z {1\over( z-|r|)^*}
 {1\over z- z'}
\eeq
in which $z,z',r $  are complex variables ($z'\neq |r|$)
and $d^2z\equiv {i\over 2} dz\wedge dz^*$.
A trick to simplify things is to use Stokes' theorem: given a one form $\Omega$
regular in the domain $\Sigma$ one has
\beq
\int_{\Sigma} \!\! d \Omega = \int_{\de \Sigma} \!\! \Omega .
\eeq
If we pose $\Omega= \omega(z,z^*) dz
+{\bar \omega}(z,z^*) dz^*$  we  can fix $\Omega $  in our case requiring:
\beq
({\de {\bar \omega}\over \de z} - {\de { \omega}\over \de z^*})
 = {i\over 2}
 {1\over (z-|r|)^*} {1\over z- z'}
.\eeq
With the choice  $ {\bar \omega}=0$ and
 $\omega=-{i\over 2}  \log (z^*-|r|)/(z-z')$
we get
\beq
I=-{i\over 2} \int_{\gamma}\!\!  \log (z^*-|r|)/(z-z') dz
,\eeq
in which the path $\gamma$ should not enclose singularities
(pole in $0$ and cut from $|r|$ that we choose along the real positive axis)
 to allow for application
of Stokes' Theorem.
We split $\gamma=\gamma_1\cup \gamma_2\cup\gamma_3\cup\gamma_4$
where (in terms of z):
\bea
& &\gamma_1\equiv \{ z / z=z'+\epsilon e^{i \phi}, \phi\in (2\pi,0)\}\non\\
& &\gamma_2\equiv \{ z / z=\rho\in[|r|, R] \;\;{\rm or} \;\;z=e^{2\pi i}\rho,
\rho\in(R,|r|)\}\non\\
& &\gamma_3\equiv \{z/z=R e^{i \phi} ,\phi\in(0,2\pi) \}
\eea
(notice the orientations of the intervals).
The contribution of the path $\gamma_4$ turning around $|r|$ and
connecting the two branches of $\gamma_2$ is easily seen to be zero.

Performing easy integrations we get the desired result:
\beq
I\sim - \pi \log\left( {|z'-|r||^2 \over R^2}\right)
+O({|r|\over R},{|z'| \over R})
\eeq

All the integrals  in Eqs.(\ref{results}) can be computed at relevant order
 by (repeated) application of Stokes' Theorem.

\section{Computation of the Hadamard finite part}\label{finitepart}
In one dimension
Hadamard finite part \cite{schwartz} of a non locally summable
(in $x=0$) function
\beq
g(x)=\sum_\nu {A_\nu\over x^{\lambda_\nu}} + {A_1\over x}+h(x)
\eeq
(in which $\lambda_\nu$ is a finite set of complex indices with
 $ \Re \lambda_\nu \ge 1$ but different from $1$ and $h(x)$ is regular)
is
\beq
Fp \int_0^b\!\!  g(x) dx \equiv \lim_{\epsilon \to 0}
\left\{
\int_\epsilon^b \!\!  g(x) -I(\epsilon) \right\}
\eeq
in which the infinite part is defined as
\beq
I(\epsilon )\equiv  \sum_\nu {A_\nu\over\lambda_\nu-1 }
({1\over \epsilon})^{\lambda_\nu-1} + {A_1}\log {1\over \epsilon}
.\eeq
In this case we obtain
\beq
Fp \int_0^b \!\! g(x) dx =- \sum_\nu {A_\nu\over\lambda_\nu-1 }
({1\over b})^{\lambda_\nu-1} -{A_1}\log {1\over b} +\int_0^b\!\! h(x) dx.
\eeq

One can extend
a locally non summable function as a tempered distribution by taking the
Hadamard finite part of the
integral of the function times the test function.
For our purposes we are involved only in radially symmetric
functions and this case is a trivial generalization of one dimensional
 situation.
Extension to higher number of dimensions can be found in \cite{schwartz} and
\cite{gelfand}.
In particular in \cite{gelfand} is given a detailed treatement of
 distributional extensions of homogeneous functions of many variables that
 could be useful in the treatment of conformal field theories singularities.
Notice also that Hadamard finite part, subtracting only divergent part is
a Minimal Subtraction in the sense explained in Section \ref{ipotesi}.

We compute now the integral
\beq
I=\int\!\! d^2 r \fp {m^2 \over 4 \pi^2} (K_1^2(mr) -K_0^2 (mr) )
\eeq
The
 integral of this distribution is easily done by
taking the finite part at $\mu r=0$ i.e. by subtracting the divergent part in
 $\mu r=0$ (we added a renormalization point scale $\mu $ to keep correct
dimensions):
\bea \non
& &
\fp\int_{\epsilon \le |r|}\!\!
 d^2r   {m^2 \over 4 \pi^2} (K_1^2(mr) -K_0^2 (mr) )
={1 \over 2\pi} \fp\int_{m\epsilon}^{\infty}\!\! dx \;x (K_1^2(x) -K_0^2 (x))\\
& &={1 \over 2 \pi}
\fp ({x^2\over 2}(2 K_1^2(x) -K_0(x)^2 -K_0(x)K_2(x)))_{m\epsilon}^{\infty}
\non\\
& &= -{1 \over 2\pi} \fp (\log ({m\epsilon\over 2}) +1 -\psi(1))\non\\
& &= -{1 \over 2 \pi}  (\log ({m\over 2\mu}) +1 -\psi(1)) \label{massiveint}
\eea
where the properties of the $K_\nu(z)$ functions (in particular their
 behavior
in $z=0$ and in $z=\infty$) have been used, see for instance \cite{gr}.

\end{document}